\documentclass[12pt]{article}
\usepackage{epsfig}

\def\beq{\begin{equation}}
\def\eeq#1{\label{#1}\end{equation}}
\def\eeqn{\end{equation}}

\def\beqa{\begin{eqnarray}}
\def\eeqa#1{\label{#1}\end{eqnarray}}
\def\eeqan{\end{eqnarray}}

\let\bar=\overbar

\def\Dslash{\not{\hbox{\kern-4pt $D$}}}
\def\dslash{\not{\hbox{\kern-2pt $\del$}}}

\def\msb{{\bar{\ssstyle M \kern -1pt S}}}

\newcommand{\nn}{\nonumber}
\def\Title#1{\begin{center} {\Large {\bf #1} } \end{center}}

\begin{document}

\Title{Charming Penguins Saga\footnote{Talk given by M. Ciuchini.}}

\bigskip\bigskip

\begin{raggedright}

{\it M. Ciuchini$^{\,\star}$, E. Franco$^{\,\ddagger}$, G. Martinelli$^{\,\ddagger}$,
M. Pierini$^{\,\ddagger}$ and L. Silvestrini$^{\,\ddagger}$\index{Ciuchini, M.}
\index{Franco, E.}\index{Martinelli, G.}\index{Pierini, M.}
\index{Silvestrini, L.}\\
$^\star$ Dip. di Fisica, Universit\`a di Roma Tre and I.N.F.N. Sezione di Roma III\\
~~Via della Vasca Navale 84\\
~~I-00146 Roma, ITALY\\
$^\ddagger$ Dip. di Fisica, Universit\`a  di Roma ``La Sapienza" and I.N.F.N. Sezione di Roma\\
~~P.le A. Moro 2\\
~~I-00185 Roma, ITALY}
\bigskip\bigskip
\end{raggedright}

\section{Introduction}
``Charming penguins" started back in 1997, coming out of a study
aimed to evaluate hadronic effects in non-leptonic two-body $B$
decays. During the years, several episodes added to the
saga:
\begin{enumerate}
\item {\it The penguin menace}~\cite{Ciuchini:1997hb,Ciuchini:1997rj}.
A lattice-inspired Wick-contraction parametrization of hadronic amplitudes
was introduced and the
observation was
put forward that non-factorizable penguin contractions of current-current
operators containing two $c$ quarks (the charming penguins) could give
large contributions in some $B$ decay channels, notably $B\to K\pi$
(a similar idea was already present in ref.~\cite{Colangelo:1989gi}).
\item {\it The neat hack of the clones}~\cite{Buras:1998ra}. The original Wick-contraction
parametrization was modified by Buras and Silvestrini. The
hadronic matrix elements were expressed in terms of new
renormalization-group invariant parameters given by suitable combinations
of the old ones. Many $B$ decay channels
were classified according to the new parametrization. Charming
penguins became a more complex object, containing further contractions
(annihilations, penguin contractions of penguin operators) in addition to
the original one.
\item {\it A new hope}~\cite{Beneke:1999br,Beneke:2000ry}.
The one-loop proof that factorization of
hadronic matrix elements holds in the limit $m_b\to\infty$ puts
phenomenological approaches based on factorization on a firmer
theoretical ground (other theoretical approaches to factorization in the infinite
mass limit were already developed, although not at the same level of
accuracy~\cite{Li:fpcp}). {\it In this limit}, non-factorizable corrections
were shown to be computable using perturbation theory. Perturbative
penguins turned out to give in general small contributions. Charming
penguins seemed at loss.
\item {\it Charming penguins strike back}~\cite{Ciuchini:2001gv}.
Using $B\to K\pi$ data,
it was shown that the parameter accounting for charming penguins has
the expected size of a $\Lambda_{QCD}/m_b$ correction. Therefore, a
sizable non-perturbative effect of charming penguins is not in
disagreement with the results on factorization obtained in the infinite mass limit.  In addition, it is preferred by the data.
\item {\it The return of factorization}~\cite{Beneke:2001ev,Beneke:fpcp}.
While everybody agrees that power-suppressed terms are in general non-perturbative and
non-factorizable, it was argued that still the bulk of the $\Lambda_{QCD}/m_b$ corrections can be either factorized or, failing that,
accounted for by few parameters (this framework is called {\it improved} QCD factorization). In addition, these parameters, once
properly defined, are claimed to have negligible effects on $B\to K\pi$ branching ratios. Under these assumptions, which were
shown to be compatible with the present data, these branching ratios can be used to extract the CKM angle $\gamma$.
\end{enumerate}
Is the saga arrived to its end? Theoretically, it is not clear whether
a non-perturbative contribution such as charming penguins is large or small. A recent
calculation using renormalons found no sign of it~\cite{Neubert:2002ix}, while, on the
contrary, it is present and effective in other approaches~\cite{Isola:2001bn}. The $B\to K\pi$ data certainly call for power-suppressed
terms and charming penguins are able to provide what is needed. Other approaches, such as the popular {\it improved} QCD
factorization, are also compatible with the data, but none is able to make predictions based only on the theory, due to the
presence of phenomenological parameters. Indeed, the presence of these parameters makes us very skeptical about the possibility
of extracting the CKM angle $\gamma$ from the measurement of the $B\to K\pi$ branching ratios.

\section{Charming penguins at work}
In this section we collect the main formulae for the amplitudes of
$B\to K\pi,\pi\pi$, introducing the parametrization used in the
analysis. We refer the reader to the literature for any detail
on the origin and the properties of these
parameters~\cite{Ciuchini:1997hb,Ciuchini:1997rj,Buras:1998ra,Ciuchini:2001gv}. From ref.~\cite{Buras:1998ra}, one
reads
\begin{eqnarray}
A(B_d\to K^+\pi^-)&=& \frac{G_F}{\sqrt{2}}\Big(\lambda_t^s P_1-\lambda_u^s (E_1-P_1^{GIM})\Big)\nn\\
A(B^+\to K^+\pi^0)&=&\frac{G_F}{2}\Big(\lambda_t^s P_1-\lambda_u^s (E_1+E_2-P_1^{GIM}+A_1)\Big)+\Delta A\nn\\
A(B^+\to K^0\pi^+)&=& \frac{G_F}{\sqrt{2}}\Big(-\lambda_t^s P_1+\lambda_u^s (A_1-P_1^{GIM})\Big)+\Delta A\nn\\
A(B_d\to K^0\pi^0)&=& \frac{G_F}{2}\Big(-\lambda_t^s P_1-\lambda_u^s (E_2+P_1^{GIM})\Big)+\Delta A\\
A(B_d\to \pi^+\pi^-)&=&\frac{G_F}{\sqrt{2}}\Big(\lambda_t^d (P_1+P_3)-\lambda_u^d (E_1+A_2-P_1^{GIM})-P_3^{GIM}\Big)\nn\\
A(B_d\to \pi^+\pi^0)&=& \frac{G_F}{2}\Big(-\lambda_u^d(E_1+E_2)\Big)+\Delta A\nn\\
A(B_d\to \pi^0\pi^0)&=&\frac{G_F}{2}\Big(-\lambda_t^s (P_1+P_3)-\lambda_u^s (E_2+P_1^{GIM}+P_3^{GIM}-A_2)\Big)+\Delta A\nn\,,
\end{eqnarray}
where $\lambda_{q^\prime}^q=V_{q^\prime q}V_{q^\prime b}^*$. Neglecting the $A_i$, these parameters can
be rewritten as
\begin{eqnarray}
&&E_1=a_1^c A_{\pi K}\,,~E_2=a_2^c A_{K\pi}\,,~A_1=A_2=0\,,\nn\\
&&P_1=a_4^c  A_{\pi K}+\tilde P_1\,,~P_1^{GIM}=(a_4^c-a_4^u)  A_{\pi K}+\tilde P_1^{GIM}\,.
\label{eq:amps}
\end{eqnarray}
The terms proportional to $a_i^q$ gives the parameters computed in the limit $m_b\to\infty$ using QCD
factorization. Their definition, together with those of $A_{\pi K}$, $A_{K\pi}$, etc., can be found for instance in
ref.~\cite{Beneke:2001ev}, although power-suppressed terms included there, proportional to the chiral factors $r^\chi_{K,\pi}$,
should be discarded in eqs.~(\ref{eq:amps}). In our case, in fact, terms of $O(\Lambda_{QCD}/m_b)$
are accounted for by two phenomenological parameters: the charming-penguin parameter
$\tilde P_1$ and the GIM-penguin parameter $\tilde P_1^{GIM}$ . In $B\to K\pi$  there are no other contributions,
once flavour $SU(2)$ symmetry is used and few other doubly Cabibbo-suppressed terms, including corrections to
emission parametes $E_1$ and $E_2$, some annihilations ($A_1$) and the Zweig-suppressed contactions ($\Delta A$), are
neglected~\cite{Buras:1998ra}. On the contrary, further power-suppressed terms ($A_2$, $P_3$, $P_3^{GIM}$) enter the
$B\to \pi\pi$ amplitudes, all with the same power of the Cabibbo angle. Therefore, these modes are subject to a larger uncertainty
than the $B\to K\pi$ ones.

\begin{table}[b]
\begin{center}
%\begin{tabular}{c|ccc|c}
%Mode  & CLEO & $BaBar$ & Belle & Average \\
%    & $\cal B$ ($10^{-6}$) & $\cal B$ ($10^{-6}$) & $\cal B$ ($10^{-6}$) & $\cal B$ ($10^{-6}$) \\
%\hline
%$\pi^+\pi^-$ & $4.3^{+1.6}_{-1.5}\pm 0.5$ & $5.4\pm 0.7\pm 0.4$ & $5.1\pm 1.1\pm 0.4$ & $5.2 \pm 0.6$\\
% $\pi^+\pi^0$ & $5.4^{+2.1}_{-2.0}\pm1.5\,(<13)$ & $4.1^{+1.1+0.8}_{-1.0-0.7}$  & $7.0 \pm 2.2 \pm 0.8$ & $4.9\pm 1.1$\\
%$\pi^0\pi^0$ & $< 5.2$ & $< 3.4$ & $2.9\! \pm\! 1.5 \!\pm\! 0.6\,(< 5.6)$ \\
%\hline
%$K^+\pi^-$   & $17.2^{+2.5}_{-2.4} \pm 1.2$ & $17.8 \pm 1.1 \pm 0.8$ & $21.8 \pm 1.8 \pm 1.5$ & $18.6 \pm 1.1$ \\
% $K^+\pi^0$   & $11.6^{+3.0+1.4}_{-2.7-1.3}$ & $11.1^{+1.3}_{-1.2} \pm 1.0$ & $12.5 \pm 2.4 \pm 1.2$ & $11.5 \pm 1.3$ \\
% $K^0\pi^+$   & $18.2^{+4.6}_{-4.0}\pm 1.6$ & $17.5^{+1.8}_{-1.7} \pm 1.3$ & $18.8 \pm 3.0 \pm 1.5$ & $17.9 \pm 1.7$ \\
% $K^0\pi^0$   & $14.6^{+5.9+2.4}_{-5.1-3.3}$ & $8.2^{+3.1}_{-2.7}\pm 1.2$ & $7.7 \pm 3.2 \pm 1.6$ & $8.9 \pm 2.3$ \\
%\hline
%\end{tabular}
\begin{tabular}{ccccc}
$\vert V_{cb}\vert\!\times\! 10^{3}$ & $\vert V_{ub}\vert\!\times\! 10^{3}$ & $\hat B_K$ & $f_{B_d}\sqrt{B_d}$ (MeV) & $\xi$\\
\hline
$40.9\!\pm\! 1.0$&$3.70\!\pm\! 0.42$ & $0.86\!\pm\! 0.06\!\pm\!  0.14$& $230\!\pm\! 30\!\pm\! 15$ &
$1.16\!\pm\!0.03\!\pm\!0.04$\\[4pt]
$F_K(M_K^2)$ & ${\cal B}(K^+\pi^-)\!\times\! 10^6  $ &  ${\cal B}(K^+\pi^0)\!\times\! 10^6  $ & ${\cal B}(K^0\pi^+)\!\times\! 10^6  $&
 ${\cal B}(K^0\pi^0)\!\times\! 10^6  $\\
\hline
$0.32\pm0.12$ & $18.6 \pm 1.1$  & $11.5 \pm 1.3$  & $17.9 \pm 1.7$ & $8.9 \pm 2.3$\\[4pt]
$F_\pi(M_\pi^2)$ & ${\cal B}(\pi^+\pi^-)\!\times\! 10^6 $ &  ${\cal B}(\pi^+\pi^0)\!\times\! 10^6  $ & ${\cal B}(\pi^0\pi^0)\!\times\!
 10^6 $ & \\
\hline
$0.27\pm0.08$&$5.2 \pm 0.6$ &$4.9\pm 1.1$ & $<\! 3.4\,${\small\it BaBar} &\\[4pt]
\end{tabular}
\caption{Values of the input parameters used in our analysis. The CP-averaged branching ratios ${\cal B}$ are taken
from ref.~\cite{Patterson:fpcp}.}
\label{tab:expbr}
\end{center}
\end{table}

Using the inputs collected in Table~\ref{tab:expbr},
we fit the value of the complex parameter $\tilde P_1=(0.13\pm 0.02)\, e^{\pm i (114\pm 35)^o}$
in units of $f_\pi F_\pi(M_\pi) $. Notice that
the sign of the phase is practically not constrained by the data. This result is
almost independent of the inputs used for the CKM parameters $\rho$ and $\eta$, namely whether
these parameters are taken from the usual unitarity triangle analysis (UTA)~\cite{Ciuchini:2000de,Ciuchini:2001zf} or only
the constraint from $\vert V_{ub}/V_{cb}\vert$ is used.

For the sake of simplicity, we also neglect here the contribution of $\tilde P_1^{GIM}$. The
$B\to K\pi$ data do not constrain this parameter very effectively, since its contribution
is doubly Cabibbo suppressed with respect to $\tilde P_1$. The remaining $\pi^+\pi^-$ mode
alone is not sufficient to fully determine the complex parameter $\tilde P_1^{GIM}$. It
is interesting, however, to notice that the GIM-penguin contribution is potentially
able to enhance the ${\cal B}(B\to\pi^0\pi^0)$ up to few $\times 10^{-6}$~\cite{Ciuchini:2001gv}.

\begin{table}[t]
\begin{center}
\begin{tabular}{c|cc|cc}
  Mode  & \multicolumn{2}{c}{UTA} & \multicolumn{2}{c}{$\vert V_{ub}/V_{cb}\vert$} \\
        & $\cal B$ ($10^{-6}$) & $\vert{\cal A}_{CP}\vert$ &  $\cal B$ ($10^{-6}$) & $\vert{\cal A}_{CP}\vert$\\
\hline
 $\pi^+\pi^-$ & $8.9 \pm 3.3$ & $0.37\pm0.17$    & $8.7 \pm 3.6$ & $0.39\pm0.20$    \\
 $\pi^+\pi^0$ & $5.4 \pm 2.1$ & -- & $5.5 \pm 2.2$ & --    \\
 $\pi^0\pi^0$ & $0.44\pm 0.13$ & $0.61\pm0.26$   & $0.69 \pm 0.27$ & $0.45\pm0.27$  \\
\hline
 $K^+\pi^-$   & $18.4 \pm 1.0$ & $0.21\pm0.10$   & $18.8 \pm 1.0$ & $0.21\pm0.12$   \\
 $K^+\pi^0$   & $10.3 \pm 0.9$ & $0.22\pm0.11$   & $10.7 \pm 1.0$ & $0.22\pm0.13$   \\
 $K^0\pi^+$   & $19.3 \pm 1.2$ & $0.00\pm0.00$   & $18.1 \pm 1.5$ & $0.00\pm0.00$   \\
 $K^0\pi^0$   &  $8.7 \pm 0.8$ & $0.04\pm0.02$   &  $8.2 \pm 1.2$ & $0.04\pm0.03$   \\
\hline
\end{tabular}
\caption{Predictions for CP-averaged branching ratios $\cal B$ and absolute value of the
CP asymmetries $\vert {\cal A}_{CP}\vert$. The left (right) columns show
results obtained using constraints on the CKM parameters $\rho$ and $\eta$ obtained from the UTA
(the measurement of $\vert V_{ub}/V_{cb}\vert$). The last four channels are those used for fitting the charming penguin
parameter $\tilde P_1$.}
\label{tab:results}
\end{center}
\end{table}
%%%%%%%%%%%%%%%%%%%%%%%%%%%%%%%%%%%%%%%%%%%%%%%%%%%%%%%%%%%%%%%%%%%%%%%%%%%

Table~\ref{tab:results} shows the predicted values of the CP-averaged
branching ratios ${\cal B}$ and the absolute value of the CP-asymmetries $\vert{\cal A}_{CP}\vert$ for
the $B\to K\pi$ and $B\to\pi\pi$ modes, since the data are not able to fix the sign of asymmetries. 
Charming penguins are able to reproduce the $K\pi$ data and are also consistent with the only $\pi\pi$
mode measured so far. It is interesting to notice that the latest measurements improve
the consistency, for a comparison see refs.~\cite{Ciuchini:1997rj,Ciuchini:2001gv}.

%%%%%%%%%%%%%%%%%%%%%%%%%%%%%%%%%%%%%%%%%%%%%%%%%%%%%%%%%%%%%%%%%%%%%%%%%
%%
%%   use this format to include an .eps figure into your paper
%%
%\begin{figure}[htb]
%\begin{center}
%\begin{tabular}{cc}
%\epsfig{file=p1uta.eps,height=2.5in} & \epsfig{file=phi1uta.eps,height=2.5in}
%\end{tabular}
%\caption{Plan of the magnet used in the Mesmeric studies.}
%\label{fig:magnet}
%\end{center}
%\end{figure}
%%%%%%%%%%%%%%%%%%%%%%%%%%%%%%%%%%%%%%%%%%%%%%%%%%%%%%%%%%%%%%%%%%%%%%%%%%%

\section{Remarks on the different approaches}
Since the different approaches aiming at evaluating power-suppressed terms contain phenomenological parameters, it is
natural to ask whether, after all, they are equivalent or not, even if the physical mechanism invoked to introduce the
parameters is not the same. To answer this question, it is useful to compute the parameters $\tilde P_1$ and $\tilde
P_1^{GIM}$ within {\it improved} QCD factorization. They read
\begin{equation}
\tilde P_1 = r^\chi_K a_6^c A_{\pi K}+b_3 B_{\pi K}\,,~\tilde P_1^{GIM} = r^\chi_K (a_6^c-a_6^u) A_{\pi K}\,,
\end{equation}
where the functions $a_i^q$ ($b_i$) contain the complex parameter $\rho_H$ ($\rho_A$), see ref.~\cite{Beneke:2001ev} for the
definitions. These two parameters account for chirally-enhanced terms, originating from hard-spectator interactions and annihilations
respectively, which are not computable within the {\it improved} QCD factorization.

The functional dependence of the amplitudes on the phenomenological parameters in the two approaches is different. For instance,
the GIM-penguin parameter is a pure short-distance correction in the {\it improved} QCD factorization, since
the $\rho_H$ dependence cancels out in the difference $a_6^c-a_6^u$. In practice, however, the main contribution of the
phenomenological parameters to the $B\to K\pi$ amplitudes comes from the annihilation term $b_3$, i.e. from $\rho_A$ .
This term behaves effectively as the charming-penguin parameter, enhancing the Cabibbo-favored amplitude.

Notice that a vanishing $\rho_A$ (and $\rho_H$), which turns out to be compatible with the data, does not mean that the
phenomenological contribution is negligible. In fact, the parameters are defined so that the phenomenological terms are
functions of $X_{A(H)}=(1+\rho_{A(H)})\log(m_B/\mu_h)$, where the scale $\mu_h$ is assumed to be 0.5
GeV~\cite{Beneke:2001ev}.

\section{Non-leptonic \boldmath$B$ decays and the extraction of \boldmath$\gamma$}
The presence of complex phenomenological parameters in the amplitudes makes the extraction of $\gamma$ very
problematic. We checked using the $\vert V_{ub}/V_{cb}\vert $-constrained fit that almost any value of $\gamma$ is
allowed, given the uncertainty on $\tilde P_1$. This seems a general problem which make us doubt recent claims
proposing non-leptonic $B$ decays as an effective tool for the CKM matrix determination. Even more, we think that the
combination of the contraint from $B\to K\pi$ decays on $\gamma$ with the others can even be misleading. The
reason is very simple: $\gamma$ is looked for through the effect of interefence terms in the branching ratios. The
presence of a competing amplitude with a new phase, i.e. the one containing the phenomenological parameter, makes
the extraction of $\gamma$ much more complicated. Although weak and strong phases can be disentangled in
principle, in practice we checked that not only the task is very difficult now, but the situation improves slowly as
data become more accurate, even when the CP asymmetries will be measured.

Concerning various analyses based on the {\it improved} QCD factorization claiming to
find a ``large'' value of $\gamma\sim 90^o$, we just notice that, as far as we know, they all assume the
bound $\vert\rho_A\vert <1$, suggested in ref.~\cite{Beneke:2001ev} as a theoretical prejudice and
supported by the observation that even $\vert\rho_A\vert = 0$ produces a good fit to ${\cal B}(B\to K\pi)$.
A better fit, however, can be obtained letting $\vert\rho_A\vert$ take values up to about $3$. As shown in
ref.~\cite{Ciuchini:2001zf}, by doing so, the contribution of the constraint from non-leptonic $B$ decays to a global fit
of $\gamma$ becomes totally negligible. In other words, for $\vert\rho_A\vert\sim 3$, the annihilation amplitude
containing $\rho_A$ becomes competitive with the others, improving the fit to the ${\cal B}$s on the one hand and
weakening the predictivity on $\gamma$ on the other.

\end{document}